\documentclass[12pt]{article}
\usepackage{amsmath,amssymb,amsthm,color}
\usepackage{graphicx}
\usepackage{cite}
\usepackage{epsfig}
\renewcommand{\baselinestretch}{2}
\begin{document}
%
\title{H-Si bonding-induced unusual electronic properties of silicene: a method to identify hydrogen concentration \\}
\author{
\small Shih-Yang Lin$^{1}$, Shen-Lin Chang$^{1,2,\star}$, Ngoc Thanh Thuy Tran$^{1,\dag}$, Po-Hua Yang$^{3}$, and Ming-Fa Lin$^{1,\ddag}$ $$\\
\small  $^1$Department of Physics, National Cheng-Kung University, Tainan 701, Taiwan\\
\small  $^2$Department of Physics, National Sun Yat-Sen University, Taiwan\\
\small  $^3$National Center for High-Performance Computing, Taiwan, 710 Tainan, Taiwan\\
 }
\renewcommand{\baselinestretch}{1}
\maketitle

\renewcommand{\baselinestretch}{1.4}
\begin{abstract}

Hydrogenated silicenes possess peculiar properties owing to the strong H-Si bonds, as revealed by an investigation using first principles calculations. The various charge distributions, bond lengths, energy bands, and densities of states strongly depend on different hydrogen configurations and concentrations. The competition of strong H-Si bondings and weak sp$^3$ hybridization dominate the electronic properties. Chair configurations belong to semiconductors, while the top configurations show a nearly dispersionless energy band at the Fermi level. Both two systems display H-related partially flat bands at middle energy, and recovery of low-lying $\pi$ bands during the reduction of concentration. Their densities of states exhibit prominent peaks at middle energy, and the top systems have a delta-funtion-like peak at $E=0$. The intensity of these peaks are gradually weakened as the concentration decreases, providing an effective method to identify the H-concentration in scanning tunneling spectroscopy experiments.

\vskip 1.0 truecm
\par\noindent

\vskip1.0 truecm

\par\noindent  * Corresponding author.
{~ Tel:~ +886-6-275-7575.}\\~{{\it E-mail addresses}: lccs38@hotmail.com (S.L. Chang), thuytran74vn@gmail.com (N.T.T. Tran), mflin@mail.ncku.edu.tw (M.F. Lin)}
\end{abstract}

\pagebreak
\renewcommand{\baselinestretch}{2}
\newpage

{\bf 1. Introduction}
\vskip 0.3 truecm

Two-dimensional materials composed by group IV elements have attracted considerable attention in the fields of chemistry, material science and physics,$^{1-6}$ mainly owing to their planar structure with sp$^2$ or sp$^3$ bondings formed by four valence electrons. This has provided the motivation to conduct studies on their electronic,$^{7-12}$ optical,$^{13-16}$ and transport$^{17-20}$ properties. The gap opening in the electronic properties is an important issue for its possible use in future nanoelectronic devices.$^{21-24}$ Several experimental methods have been realized, one of which, the chemical functionalization,$^{25-31}$ is particularly promising. Recently, the successful hydrogenation of graphene has created a novel material known as graphane with a large band gap,$^{32-35}$ which has been verified by angle-resolved photoemission spectroscopy (ARPES).$^{33-35}$ During the synthesis process, its band structure can be significantly altered and controlled through reversible hydrogenation,$^{32}$ while the concentration is very difficult to estimate. However, graphene faces many challenges in applications, such as toxicity, difficulty in processing, issues to integrate it with current silicon-based electronic technology, and the availability of large-area sheets. This work investigates in detail how the unique H-Si bonds with special charge distributions induce the feature-rich geometric and electronic properties in hydrogenated silicene.

As the counterpart of graphene, 2D hexagonal lattices of Si, namely silicene, and Si nanoribbons have been fabricated by the chemical vapor deposition process on a silver substrate$^{3,36-38}$ and Iridium substrate.$^{39}$ Other chemical exfoliation methods have succeeded in creating suspended silicon monolayer sheets doped with magnesium and covered with organic groups.$^{40-42}$ These experimental realizations of providing suitable environments for single-sided and double-sided chemical functionalizations further tune the electronic properties of silicene drastically. Quite different from graphene or other carbon-based nanostructures, silicene with its diverse electronic properties could enable the electronic industry to produce fast nanoscale electronics without the need to retool.

Unlike graphene, silicene is not stable as a perfectly planar sheet. Because the sp$^3$ hybridization is more stable in silicene than the sp$^2$ hybridization, it is energetically favorable as a low buckled structure, which has been confirmed by DFT studies and phonon dispersion calculations.$^{43}$ Recently, the electronic structure of silicene has been revealed to be that of a zero-gap semiconductor by ARPES and scanning tunneling spectroscopy (STS) measurements.$^{3,5,44}$ Other theoretical works exploring silicene sheets passivated with hydrogen show the emergence of an indirect middle band gap.$^{45,46}$ These results suggest that the electronic properties of hydrogenated silicene can be altered by a competition between sp$^3$ hybridization and hydrogenation. However, since silicene sheets have only been recently realized, the effects of various hydrogen concentrations with different configurations on this new material have yet to be thoroughly explored.

In this paper, how to create the diversified electronic properties via competition between the H-Si bonds and the Si-Si $\sigma$ and $\pi$ bonds is investigated by first principle calculations. The double-sided (chair configuration in Fig. 1(a)) and the single-sided hydrogenations (top configuration in Fig. 1(b)) are taken into consideration to reveal that there are two basic types of electronic structures, which could be further modulated by various hydrogen concentrations. This work shows many kinds of energy dispersions, including the presence or the absence of a Dirac cone, the H-Si-bond-induced partially flat bands, and the nearly dispersionless band at the Fermi level induced by broken mirror symmetry. All the predicted band structures are further reflected in various feature-rich  densities of states (DOS), such as strong H-concentration-related peaks. The effect of hydrogenation on band structures and DOS could be understood by a thorough analysis of the charge distribution of $\pi$ and $\sigma$ bondings. The predictions of the band gaps and dispersions, and critical characteristics in DOS could be further identified by ARPES and STS,$^{3,5,47,48}$ respectively, with the latter being very useful in determining the hydrogen concentration due to its relation with the peak intensities. Hopefully, these rich fundamental features in hydrogenated silicenes structures can promote potential applications in energy materials or electronic devices.

\vskip 0.6 truecm
\par\noindent
{\bf 2. Methods }
\vskip 0.3 truecm

Our first-principle calculations are based on the density functional theory (DFT) implemented by the Vienna \emph{ab initio} simulation package (VASP).$^{50}$ The local density approximation (LDA) is chosen for the DFT calculations, and the projector augmented wave (PAW) is utilized to describe the electron-ion interactions. A vacuum space of 10 angstroms is inserted between periodic images to avoid interactions, and the cutoff energies of the wave function expanded by plane waves were chosen to be 500 eV. For calculations of the electronic properties and the optimal geometric structures, the first Brillouin zones are sampled by $100\times100\times1$ and $12\times12\times1$ \emph{k}-points via the Monkhorst-Pack scheme. The convergence of the Helmann-Feymann force is set to 0.01 eV ${\AA}^{-1}$.

\vskip 0.6 truecm
\par\noindent
{\bf 3. Results and Discussion}
\vskip 0.3 truecm

There exist specific relationships between the optimal geometric structures of hydrogenated silicene and various hydrogen configurations. Two different atomic configurations of fully hydrogenated silicene are taken into account. The chair-like configuration, where H atoms alternatively appear on both sides of silicene plane is shown in Fig. 1(a). The H atoms are set to be in a uniform distribution with each H atom being separated from its six nearest H atoms by the same distance (Fig. 1(c)), and the back-site H atoms (green circles) are located at the middle of three front-site H atoms (red circles). The hydrogen concentration is reduced from 100 \% to 25 \%, and then 6.3 \%, in which the hydrogen-silicon ratios correspond to 2 : 2, 2 : 8, and 2 : 32, respectively (Fig. 1(c)). For the top configuration, hydrogen atoms only appear on a single side of the plane (Fig. 1(b)), i.e., all the back-site H atoms are removed, so the H-concentration is half of the chair-like one. These two types are revealed to have similar geometric deformations; however, they may lead to diverse electronic properties, mainly owing to the breaking of the mirror symmetry.

The main characteristics of geometric structures, including the H-Si-Si angles, Si-Si bond lengths, and H-Si bond lengths, are dominated by the H-concentration, as shown in Table 1. The H-Si-Si angles present a significant change from $107.0 ^\circ$ to $111.7 ^\circ$, as the H-concentration is reduces from 100 \% to 3.1 \%. The Si-Si bond lengths are non-uniform in hydrogenated silicene, while they are at the same value of 2.25 {\AA} in pristine silicene. The strong H-Si covalent bonds are formed by the obvious attraction of electrons from Si atoms to H (discussed later). Such charge transfer makes the nearest Si-Si bond weaker but the second-nearest Si-Si bond stronger, as view from the H atom. As a result, the bond lengths get longer and shorter, respectively, in the range of $2.31 \sim 2.33$ {\AA} and $2.23 \sim 2.24$ {\AA}. It should be noted that the planar Si-Si bond lengths are mainly determined by the $\sigma$ bondings due to the (3s, 3p$_{x}$, 3p$_{y}$) orbitals. However, the H-Si bond lengths are almost insensitive to the change in H-concentration, in which they are kept at 1.50 {\AA} because of the strong and stable covalent $\sigma$ bondings. The above-mentioned bonding configurations are closely related to the strong covalent bonds between 3p$_{z}$ and 1s orbitals, and the weak sp$^3$ hybridization of four orbitals (3s, 3p$_{x}$, 3p$_{y}$, 3p$_{z}$) with 3p$_{z}$ orbitals perpendicular to the plane (discussed later in Fig. 4(a)). These rich geometric configurations lead to the diverse electronic properties.

Hydrogenated silicene exhibits very special energy bands, being enriched by the competition of strong H-Si bondings and weak sp$^3$ hybridization. Pristine silicene, as shown in Fig. 2(a), has an almost vanished gap of $\sim 1$ meV at the K point, and its low-lying linear dispersions become parabolic bands with a saddle point at the M point. Such bands can be regarded as $\pi$ bands, mainly coming from the bonding of 3p$_{z}$ orbitals of two nearest Si atoms. The weak sp$^3$ hybridization indicates a clear separation of the $\pi$ bands and $\sigma$ bands within $\pm 2$ eV. The occupied and unoccupied $\sigma$ bands, with parabolic energy dispersions, are initiated at about $-1.2$ eV and $1$ eV, respectively. The highest occupied states (HOS) are doubly degenerate at the $\Gamma$ point, while the lowest unoccupied states (LUS) are non-degenerate at the M point. The former energy bands are closely related to the (3p$_{x}$, 3p$_{y}$) orbitals with their orbital hybridization with 3p$_{z}$ orbitals at lower energy ($\sim -2.5$ eV). The third $\sigma$ band with an initial energy at $-11.2$ eV, dominated by the 3s orbitals (not shown), gradually increases with a parabolic behavior. However, it becomes partially flat at $-3.2$ eV near the $\Gamma$ point, where the 3s orbitals hybridize with the 3p$_{z}$ orbitals (discussed later). The main features of silicene are dramatically changed after hydrogenation.

The chair-like systems exhibit semiconducting behavior with an H-concentration-dependent energy gap. Fully hydrogenated silicene does not exhibit a linear Dirac cone as a result of the strong H-Si bonding, as illustrated in Fig. 2(b). It has a large indirect band gap of $E_{g}=2.1$ eV, which is determined by the HOS and the LUS at the $\Gamma$ and the M points, respectively. All 3p$_{z}$ orbitals are saturated by 1s orbitals so that the low-lying $\pi$ bands are replaced by the $\sigma$ bands corresponding to Si atoms, with the amount of contribution indicated by the radius of red circles in Fig. 2(b). Compared to Fig. 2(a), a clear revelation is that the valence $\sigma$ bands are formed by the (3p$_{x}$, 3p$_{y}$) orbitals. On the other hand, a weakly dispersed band at about -$3.5$ eV mainly originates from one of the particular 1s orbital of two H atoms (blue circles), and the other 1s orbitals contributes equally to two partially flat bands. The former characteristic is very useful in gauging the H-concentraion. Such electronic states could be roughly regarded as localized states. The H-dependent localization behavior reflects the fact that the 1s orbital has much lower bound state energy than the other orbitals. Moreover, the hybridization of 1s and 3p$_{z}$ orbitals puts the $\pi$ bands far away from $E_{F}=0$, and thus the energy gap is determined by the $\sigma$ bands.

As the concentration decreases, the H-Si bonds no longer dominate, but compete with the effect of weak sp$^3$ hybridization to determine the band structures. The 25 \% and 6.3 \% systems, namely the intermediate states between fully hydrogenated and pristine silicene, respectively, possess $1.3$ and $0.2$ eV direct band gaps at the $\Gamma$ point, as shown in Fig. 2(c) and 2(d). Their two pairs of energy bands nearest to $E_{F}=0$ exhibit weak dispersions with narrow band widths, in which they originate from the 3p$_{z}$ orbitals of non-passivated silicon atoms. This indicates that the low-lying bands are gradually changed from $\sigma$ bands to $\pi$ bands. For the middle-energy states (-$1$ eV $-$ -$3.5$ eV), the $\sigma$ bands mainly come from both the passivated (red circles in Fig. 2(c)) and non-passivated silicon atoms. The latter contribute much to such bands in the 6.3 \% systems as a result of the higher percentage. The weakly dispersed band mainly contributed by hydrogen atoms (blue circles) moves to -$4.0$ and -$4.2$ eV for 25 \% and 6.3 \% systems, respectively. For other low concentrations, the H-dependent energy gaps, as shown in Fig. 2(e), start to fluctuat and gradually decline. As expected, the tunable direct band gap in hydrogenated silicene can be used to provide potential applications in nanoelectronic and nanophotonic devices.

In the top configuration, the destruction of the mirror symmetry induces strong localized states at the Fermi level. The 50 \% system has a nearly dispersionless energy band at $E_{F}=0$ along different directions in Fig. 3(a), which is only associated with three non-passivated silicon B atoms  (green circles) nearest to the passivated silicon A atom. Such quasi-zero-dimensional electronic states are hardly affected by the H-concentration, as indicated by the 12.5 \% and 3.1 \% systems in Figs. 3(b) and 3(c), respectively. The $\pi$ bondings between B atoms and the A atoms are severely suppressed by the large charge transfer between the latter and the H atoms. As a result, B atoms cannot form the extended $\pi$-electronic states through interactions with A atoms. It should be noted that these systems are gapless, while the conducting behavior is expected to be forbidden by the almost vanishing group velocities. The critical difference between the top and the chair configurations is the intriguing localized states. In addition, the reduction of one effective 1s orbital in top systems denies the presence of the weakly dispersed band of the chair configurations.

There exist certain similarities between the top and chair configurations, including the nucleation of $\pi$ bands at low H-concentration, the Si-Si bond-related $\sigma$ bands, and the H-Si bond-induced localized states. As the H-concentration decreases, both systems exhibit the Si-atom-contributed low-lying bands with stronger dispersions and larger energy widths, as shown by the 6.3 \% (chair) and 3.1 \% (top) systems in Figs. 2(d) and 3(c), respectively. Such bands mainly result from 3p$_{z}$ orbitals, meaning that the $\pi$ bondings are reformed among many of the non-passivated Si atoms. For the $\sigma$ bands within -$1$ $-$ -$3.5$ eV (Figs. 2(a)-2(c) and 3(a)-3(c)), the initial energy, band width, and state degeneracy are much alike between two configurations, owing to the barely influenced planar $\sigma$ bonds. Moreover, the partially flat bands related to H-Si bonds (blue and red circles) are confined in a similar energy range compared to their counterparts in the chair configuration, reflecting the low bound state energy of the H atoms. The above-mentioned similarities are hardly affected by the breaking of the mirror symmetry.

The main characteristics of the energy bands can be verified by ARPES measurements, as done for silicenes and few-layer graphenes.$^{3,47,48}$ Such experiments have been utilized to confirm the graphene-like linear energy dispersions for a silicene layer on Ag(111),$^{3}$ and also to observe the opening of a substantial energy gap in hydrogenated graphene.$^{34,35}$ As expected, in the verification of hydrogenated silicene, ARPES can directly detect the gradually recovered $\pi$ bands, H-related partially flat bands, and nearly dispersionless bands at $E_{F}=0$, which are, respectively, induced by the strengthened $\pi$ bonding between 3p$_{z}$ orbitals, the strong H-Si bonds, and the broken mirror symmetry. It is deduced that the comparison between theoretical and experimental results of band gaps and energy dispersions can provide valuable information in determining the H-concentrations.

The carrier density ($\rho$) and the variation of carrier density ($\Delta \rho$) can provide very useful information in the charge distribution in bondings of all orbitals and thus explain the dramatic change of low-lying energy bands. The former directly reveals the bond strength of Si-Si and H-Si bonds, as illustrated in Figs. 4(a)-4(d). Between two Si atoms of pristine silicene or two non-passivated Si atoms of hydrogenated silicene, $\rho$ shows a strong covalent $\sigma$ bond, as indicated by an enclosed black rectangle region in Figs. 4(a), 4(c), and 4(d). Such bonds become a bit weaker when the Si atom bonds with an H atom (gray rectangle in Figs. 4(b)-4(d)). The little change is responsible for the similarities of $\sigma$ bands between the two configurations under various H-concentrations. Away from the central line between two Si atoms, carrier density is lower, but its distribution is extended between them (purple rectangle). This is evidence for the $\pi$-electronic states, revealing the gradual recovery of $\pi$ bands at a low H-concentration. However, barely extended states, corresponding to the $\pi$-electron-depleted region, are observed between the passivated and non-passivated atoms (red rectangle in Fig. 4(d)). Such charge distribution is similar to that of isolated Si atoms and demonstrates the disappearance of $\pi$ bonding. This also induces the nearly dispersionless band in all top configurations. On the other hand, the charge density of H-Si bonds (dashed rectangle) is much stronger than those of all Si-Si bonds. The very strong bond strengths are insensitive to H-concentration in both two systems, so that the H-related energy bands are confined at about -$3.5$ eV $-$ -$5.5$ eV. It is also noted that the charge distributions can explain why only the nearest Si-Si bond lengths are changed, as discussed earlier in Table I.

In order to comprehend the charge transfers among all orbitals, the detailed variation of carrier density is clearly illustrated in Figs. 4(e)-4(h). $\Delta \rho$ is created by subtracting the carrier density of isolated silicon atoms (isolated silicon and hydrogen atoms) from that of a silicene (hydrogenated silicene). For pristine silicene, as indicated in Fig. 4(e), electrons are transferred from the blue region near Si atoms to the location at the middle of two Si atoms (yellow region within a black rectangle), forming covalent $\sigma$ bonds. Distantly away from the middle region between two Si atoms, $\rho$ also increases in the green region (purple rectangle), which verifies the creation of $\pi$ bondings. On the other hand, all the hydrogenated cases display high charge transfer from Si atoms to H atoms ($\sim 0.9$ e by using Badar charge analysis), forming strong H-Si bonds (dashed rectangle Figs. 4(f)-4(h)). This also accounts for a rather weak bonding strength of the nearest Si-Si bond around the H atom.

The density of states (DOS) directly reflects the primary characteristics of the band structures, in which the partial DOS (PDOS) can be used to fully understand the orbital-decomposed contributions. For pristine silicene, DOS is zero at $E_{F}=0$; it possesses a linear E-dependence at low energy, and then a symmetric logarithmic divergent peak at $E \sim \pm 1$ eV (black curve in Fig. 5(a)). Such features, respectively originating from the almost vanished energy gaps, isotropic linear bands, and the saddle points of $\pi$ bands, are dominated by the 3p$_{z}$ orbitals (purple curve). The middle-energy DOS (-$1$ eV $-$ -$3$ eV) reveals a shoulder (-$1.2$ eV) structure and a prominent peak (-$2.7$ eV), corresponding to the maximum point and the saddle M point of the $\sigma$ bands, respectively. They are mainly contributed by the 3p$_{x}$ and 3p$_{y}$ orbitals (green and blue curves). Also, the hybridization between 3p$_{z}$ and 3s orbitals induces an extra strong peak at -$3.2$ eV. These characteristics further illustrate that $\pi$ and $\sigma$ bands have the weak interactions at low energy.

The low energy DOS is drastically changed after hydrogenation. For the fully hydrogenated silicene in the chair configuration, DOS displays a vacant region within $E \sim \pm 1$ eV, and an extra strong peak at about -$3.5$ eV and -$4.5$ eV, that comes from the H-related weakly dispersed band and the partially flat band, respectively, as shown in Fig. 5(b). These features are mainly owed to the strong bonding of 3p$_{z}$ and 1s orbitals (purple and red curves). However, the shoulder structure at $-1.0$ eV and the prominent peak at $-2.5$ eV, constituted by the 3p$_{x}$ and 3p$_{y}$ orbitals, are very similar to that of pristine silicene as a result of the small change in the $\sigma$ band. As the H-concentration decreases, two prominent peaks are revealed at $\pm 0.6$ eV in the 25 \% system, as shown in Fig. 5(c). These peaks have widths of about $0.5$ eV, and their states are associated with $\pi$ bands that are related to 3p$_{z}$ orbitals. The middle energy states are also dominated by the 3p$_{x}$ and 3p$_{y}$ orbitals; however, the single peak structure in the 100 \% system changes into several subpeaks due to an increase of the number of subbands. In the 6.3 \% system, 3p$_{z}$ orbitals contribute most of the electronic states within $\pm 1$ eV, indicating a similar $\pi$ band width to pristine silicene. The two lower concentrations show two strong peaks between -$4.0$ $-$ -$4.5$ eV, originating mainly from the 1s orbital of the H atoms (red curve) and the 3p$_{z}$ orbitals of Si atoms (purple curve). Nevertheless, the intensities of these two peaks are roughly proportional to the H-concentration, especially for the peak closer to the Fermi level.

Common features in DOS between the top and chair systems are the strong peaks associated with the planar covalent $\sigma$ bonds and the H-related bonds. Analogies are found between a top case and its corresponding chair configuration (with double the H-concentration), including the widened low-energy $\pi$ band width (Figs. 5(d) and 5(g)), the two 3p$_{z}$-orbital-induced prominent peaks at $\pm 0.6$ eV (Figs. 5(c) and 5(f)), the 3p$_{x}$ and 3p$_{y}$ orbitals dominating middle-energy peaks (Figs. 5(b)-5(g)), and the strong peaks contributed by the H-Si bonds. Such special peaks are located in almost the same energy range, while their intensities are gradually decreased with diminishing H-concentration. On the other hand, the destruction of the mirror symmetry leads to a critical difference, namely a delta-funtion-like peak at $E_{F}$ (Figs. 5(e)-(g)). Its intensity gradually decreases, directly reflecting the percentage of the nearest non-passivated Si atoms. Both 3p$_{z}$ and 3s orbitals contribute to these peaks, as shown by the purple and red curves, respectively. This directly illustrates that the non-extended $\pi$-electrons of 3p$_{z}$ orbitals are hybridized with 3s orbitals.

The STS measurements, in which the tunneling differential conductance map of the dI/dV-V curve is proportional to DOS, can provide an accurate and efficient way to examine the H-concentration. It has been adopted to identify the Dirac point of pristine silicene.$^{5}$ Hydrogenated graphene that possesses a gap larger than $2$ eV has also been measured with this method.$^{49}$ The STS measurements could be used to determine the energy gaps and the specific DOS peaks of hydrogenated silicene, especially the strong H-bond-related peaks (red triangles in Figs. 5(b)-5(d)), and the delta-funtion-like peak at $E_{F}$ (Figs. 5(e)-5(g)). These intensity of these peaks is predicted to be strongly associated with the percentage of H atoms. The intensity of the H-related peaks can be well fit by a linear relationship $ I = 1.6 I_{0} (N/N_{0})$, where $I$ is the peak intensity and $N$ is the H-concentration, with $I_{0}$ and $N_{0}$ being the value for the highest concentration. For the delta-funtion-like peak at $E_{F}$ this relation changes into $I = 1.3 I_{0} (N/N_{0})$. Given the linear dependency of DOS peaks associated with the contributions from H atoms and the nearest non-passivated atoms, we have a feasible way to identify the hydrogen concentration.

\vskip 0.6 truecm
\par\noindent
{\bf 4. Conclusion }
\vskip 0.3 truecm


The geometric and electronic properties of hydrogenated silicene are investigated by DFT calculations. Chemical and physical properties are enriched by different hydrogen configurations and concentrations, in which the special H-Si bonds are the critical factor affecting the optimized geometric structures, charge distributions, energy bands, and DOS. The high charge transfer between 3p$_{z}$ and 1s orbitals, directly or indirectly leading to variations in the bond lengths and carrier densities, is responsible for the observed dramatic change in the electronic properties. The three major revelations of the band structures are highlighted by the absence or recovery of low-lying $\pi$ bands, the H-related partially flat bands, and the nearly dispersionless band at $E_{F}=0$. The first one is determined by the competition of strong H-Si bondings and weak sp$^3$ hybridization. The band gap mainly determined by the $\sigma$ bands is largest for the 100 \% H-concentration. As the concentration drops, the influence of the $\pi$ bands gradually rises, causing the gap size to shrink. The second one, purely originating from the strong charge bondings between H and Si atoms, is a common characteristic for both top and chair systems under various H-concentrations, while the last one, being ascribed to the destruction of mirror symmetry, is a distinguished characteristic for top configurations. As for DOS, the recovery of $\pi$-bands-induced low-lying peaks, the diminishing of H-related strong peaks, and the weakened delta-funtion-like peak at $E_{F}$ directly reflect the reduced H-concentration. The feature-rich band gaps and energy dispersions, and the strong DOS peaks, respectively, could be measured by ARPES and STS experiments, giving useful information about the hydrogen distribution and concentration. Especially, the linear dependence of the special DOS peaks provides an effective path to accurately detect the H-concentration. Hydrogenated silicene is very suitable for the application of hydrogen storage, and the tunable electronic properties might be potentially important for applications in nanoelectronic and nanophotonic devices.

\par\noindent {\bf Acknowledgments}

This work is supported by the NSC and NCTS (South) of Taiwan, under the grant No. NSC-102-2112-M-006-007-MY3. We are grateful to the National Center for High-performance Computing (NCHC) for computer time and facilities.

\newpage
\renewcommand{\baselinestretch}{0.2}

\begin{figure}[htb]
\centering\includegraphics[width=0.9\linewidth]{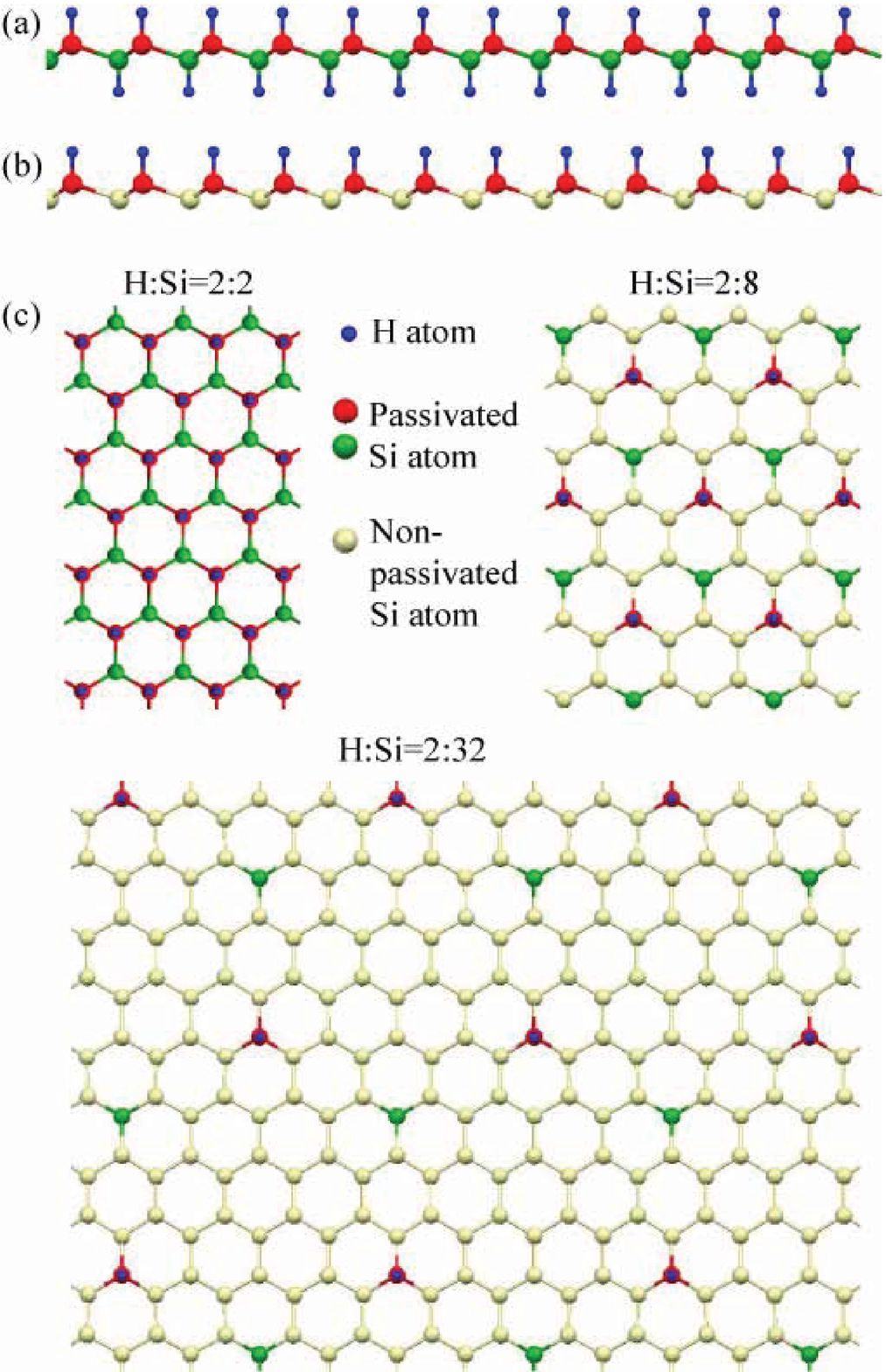}
\caption{Side view atomic configurations of hydrogenated silicenes: (a) chair configuration, where the H atoms appear on both sides of silicene plane, and (b) top configuration, where all H atoms are located on the same side of the plane. (c) top view for various concentrations, namely 2:2, 2:8, and 2:32.}
\end{figure}

\begin{figure}[htb]
\centering\includegraphics[width=0.6\linewidth]{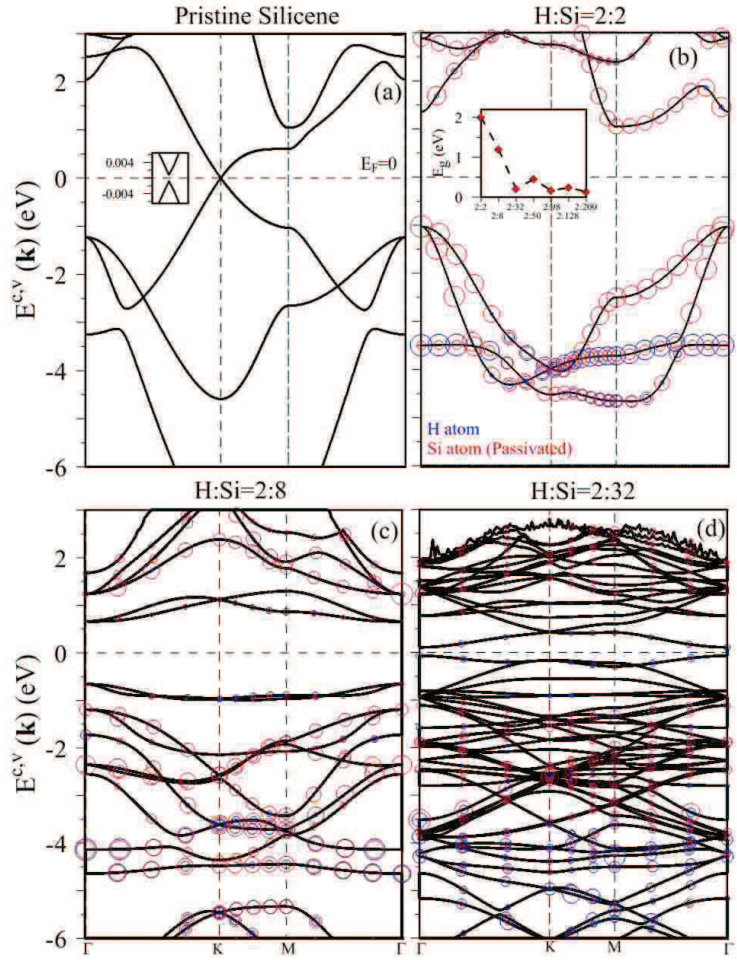}
\caption{Energy bands of (a) pristine silicence, and hydrogenated silicenes with chair configuration for (b) 2:2, (c) 2:8, and (d) 2:32 hydrogen coverages. The inset of (a) shows the low-lying energy bands, and (b) the band gaps of various concentrations.}
\end{figure}

\begin{figure}[htb]
\centering\includegraphics[width=0.9\linewidth]{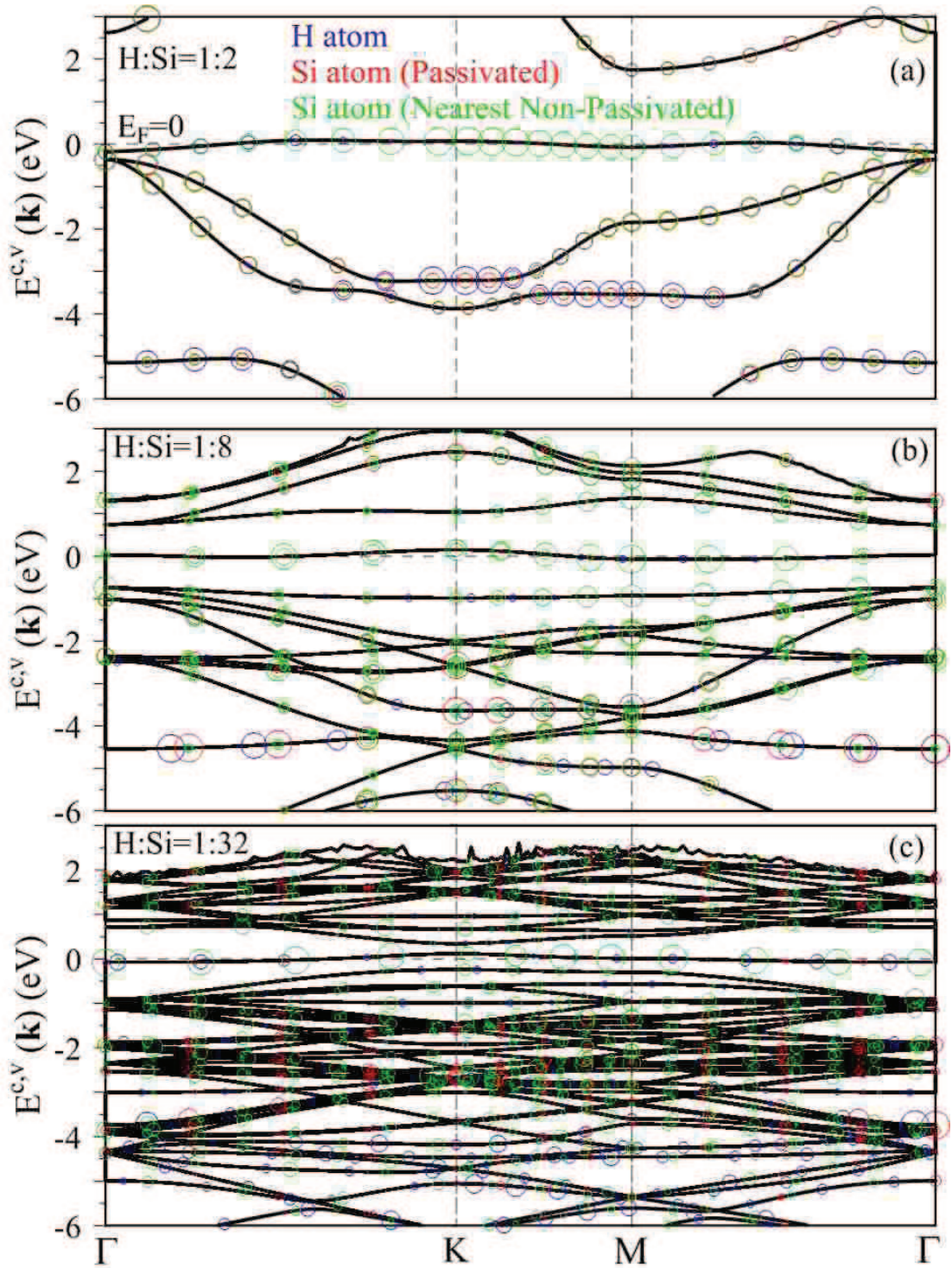}
\caption{Energy bands of top configurations for (a) 1:2, (b) 1:8, and (c) 1:32 concentrations.}
\end{figure}

\begin{figure}[htb]
\centering\includegraphics[width=0.9\linewidth]{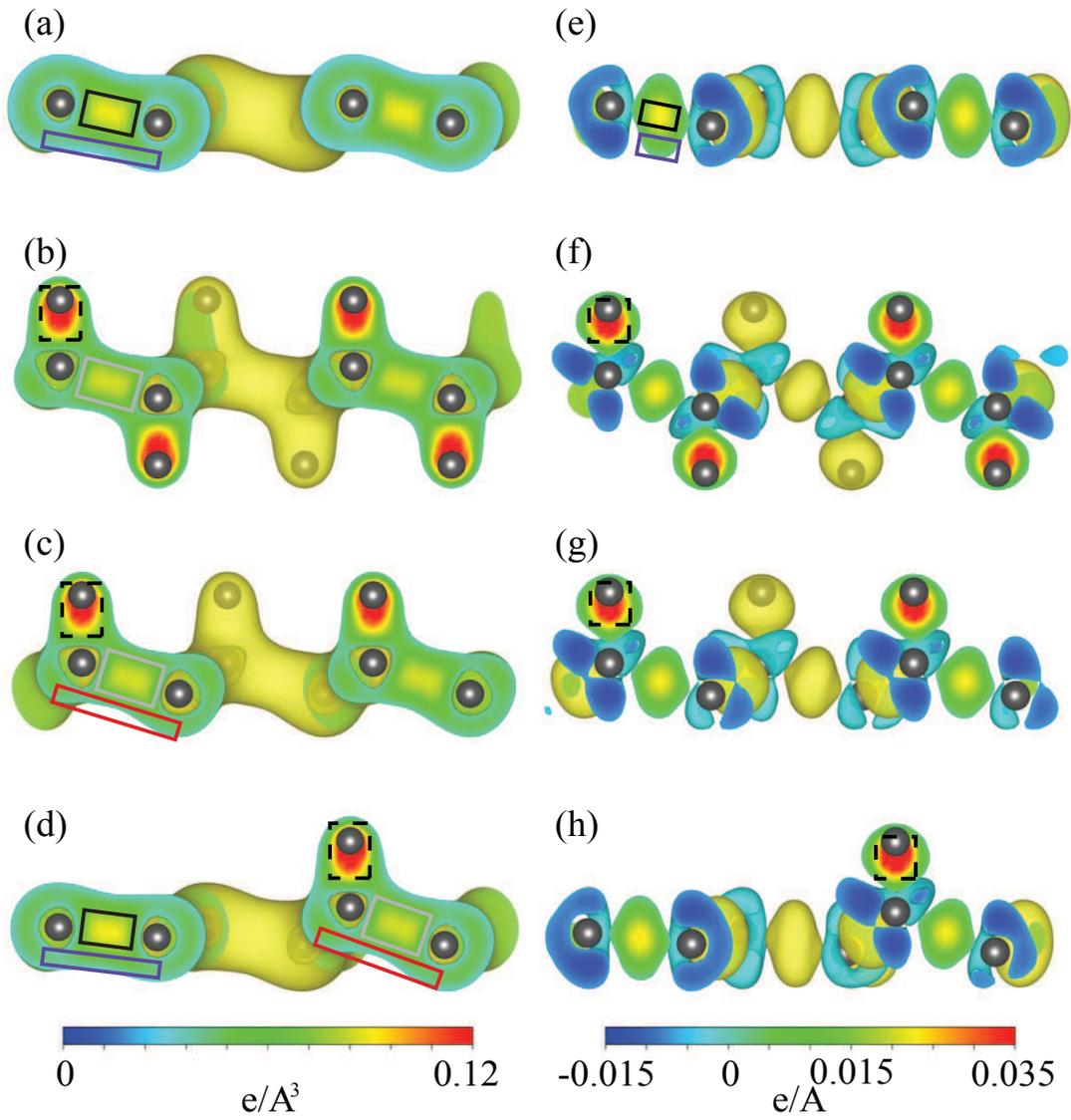}
\caption{The carrier density $\rho$ of (a) pristine silicene, and hydrogenated silicenes with (b) 2:2, (c) 1:2, and (d) 1:8 concentrations. The variation of carrier density $\Delta \rho$ of (e) pristine silicene, and hydrogenated silicenes with (f) 2:2, (g) 1:2, and (h) 1:8 concentrations.}
\end{figure}

\begin{figure}[htb]
\centering\includegraphics[width=0.9\linewidth]{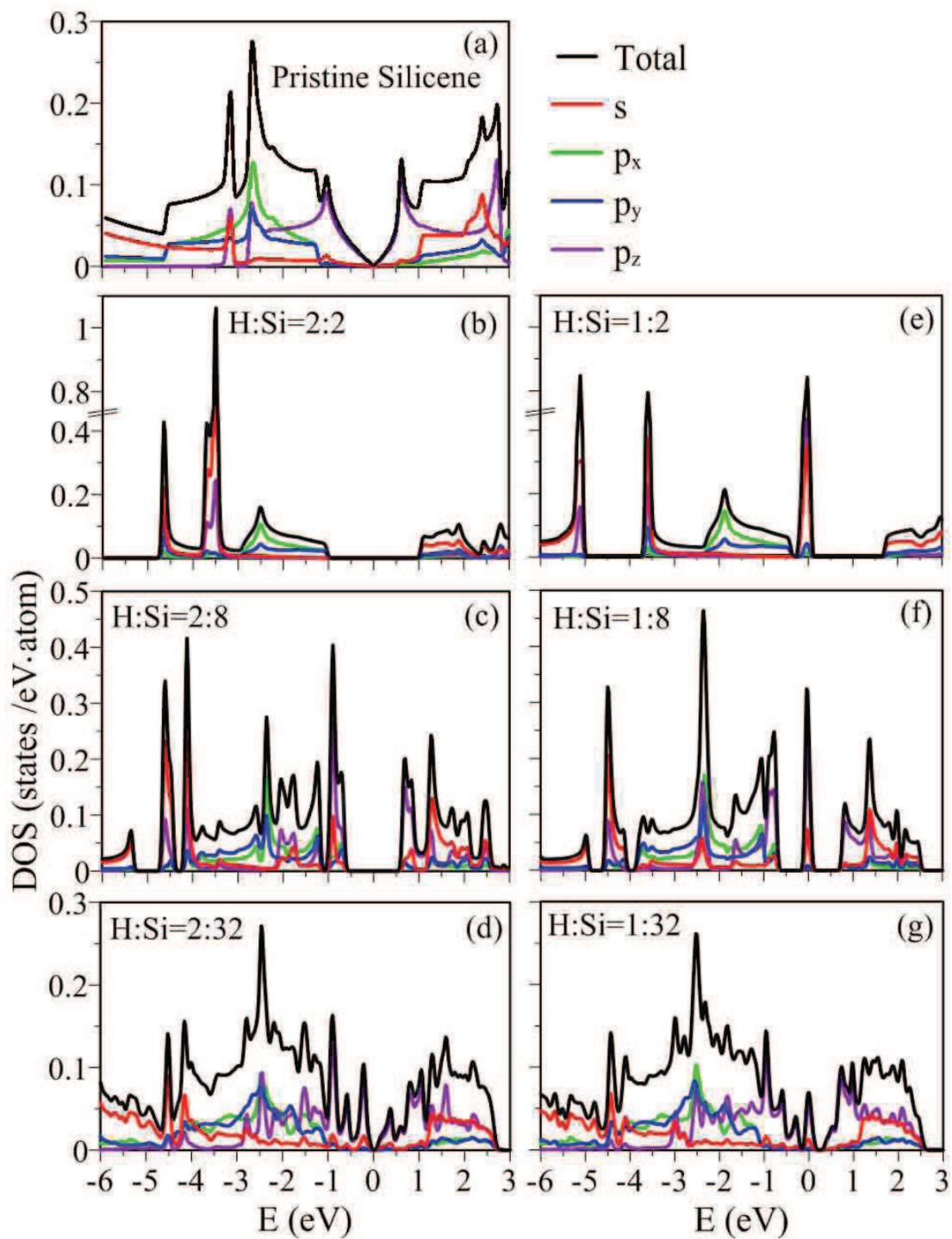}
\caption{Densities of states of (a) pristine silicenes, chair configurations with (b) 2:2, (c) 2:8, and (d) 2:32 hydrogen coverage, and top configurations with (e) 1:2, (f) 1:8, and (g) 1:32 hydrogen coverage.}
\end{figure}

\end{document}